\documentclass[aps,prd]{revtex4}
\usepackage{amsmath}
\usepackage{graphicx}

\begin{document}
\title{Simple counterterms for asymptotically AdS spacetimes \\ in Lovelock gravity}
\date{October 17, 2011}

\author{Alexandre Yale}
\email{ayale@perimeterinstitute.ca}
\affiliation{Perimeter Institute, 31 Caroline St. N., Waterloo, Ontario N2L 2Y5, Canada \\
Department of Physics \& Astronomy and Guelph-Waterloo Physics Institute,
University of Waterloo, Waterloo, Ontario N2L 3G1, Canada}

\begin{abstract}
Although gravitational actions diverge in asymptotically AdS spacetimes, boundary counterterms can be added in order to cancel out those divergences; such counterterms are known in general to third order in the Riemann tensor for the Einstein-Hilbert action.  Considering foliations of AdS with an $S^m \times H^{d-m}$ boundary, we discuss a simple algorithm which we use to generate counterterms up to sixth order in the Riemann tensor, for the Einstein-Hilbert, Gauss-Bonnet and third-order-Lovelock Lagrangians.  We also comment on other theories such as $F(R)$ gravity.
\end{abstract}
\maketitle

\newcommand{\myeq}[1]{\begin{equation} \begin{split}  #1 \end{split} \end{equation}}
\newcommand{\R}[0]{{\cal{R}}}
\newcommand{\Tr}[0]{{\text{Tr}}}
\newcommand{\bulk}[0]{{\text{bulk}}}
\newcommand{\sur}[0]{{\text{sur}}}
\newcommand{\ct}[0]{{\text{ct}}}

\section{Introduction}
An important application of the AdS / CFT correspondence \cite{Witten1998} has been the study of the thermodynamics of field theories using the tools of black hole thermodynamics.  The combination of these two ideas means that one may calculate the free energy of a strongly coupled field theory by integrating the action of a gravitational theory on a suitable background.  In particular, this has recently been used to calculate holographically the entanglement entropy of conformal field theories using black hole thermodynamics \cite{Hung2011,Casini2011,Hung2011b,Hung2011c}.
\\\\
These gravitational actions are understood to diverge in precisely the same way as their dual quantum field theories \cite{Burgess1999}.  However, the regularization techniques which deal with these divergences are quite different than their field-theoretic counterpart.  For example, in order to calculate the free energy associated with a black hole, one typically evaluates the gravitational action then subtracts a divergent piece corresponding to a background which resembles this black hole spacetime.  This technique, which goes by the name of background subtraction, can however be problematic because it is not always clear which background spacetime should be subtracted \cite{Rob1999}.  Topological black holes \cite{CCBH} are particularly problematic: for example, in the case of some hyperbolically-foliated black holes in AdS, it is unclear whether the background should be the hyperbolic foliation of AdS, which itself corresponds to a thermal state similar to the BTZ black hole, or an extremal black hole embedded in AdS, which has $T=0$ but is no longer AdS.  Moreover, in the context of a gauge/gravity duality, one would expect to find a regulating scheme for gravity which is closer to the one found in field theories, namely through the addition of counterterms.  It is important to note that these counterterms are not in competition with the Gibbons-Hawking surface terms that one must add to the action to ensure that the variational principle is well-defined: rather, it is the total action (bulk $+$ boundary) which diverges, and the surface counterterm regularizes these divergences.
\\\\
The development of such counterterms began soon after the emergence of the AdS/CFT correspondence, initially through the calculation of stress-energy tensors and as a means of renormalizing correlation functions on the gravity side of the duality \cite{Kraus1999b,Skenderis2001}, and soon after as a means of studying the thermodynamics of those topological black holes for which the background subtraction techniques failed \cite{Rob1999}.  Because this method is both physically well-motivated, through the AdS / CFT correspondence, and much simpler than other regularization techniques, it has become a standard tool in holographic calculations.  In particular, it has been key to a number of recent results, such as the derivation of thermodynamic quantities for various types of Kerr-NUT/bolt-AdS spacetimes \cite{Skenderis2005,Olea2005,Awad2000,Das2000,Mann2000} and D-branes \cite{Skenderis2006}, as well as the study of linear dilaton gravity \cite{Mann2010}.
\\\\
Through a connection with the trace anomaly in the boundary theory \cite{Skenderis1998,Skenderis1998b}, an algorithm has been developed \cite{Kraus1999} to compute these counterterms for asymptotically AdS spacetimes with an arbitrary foliation.  This algorithm consists of an iterative procedure whereby the action is varied to yield the stress-energy tensor, which is itself used to find the term that must be added to the Lagrangian to cancel its divergences to the next order.  This new improved Lagrangian is then once again integrated and varied, yielding the next order of the energy-momentum tensor, and so on.
\\\\
Despite its validity, there are obvious drawbacks to this algorithm.  Most importantly, although it allows one to easily calculate the first few counterterms, the higher-order terms have never been computed because of the rapidly increasing complexity of the algebra caused by the need to vary the action with counterterms, as pointed out in \cite{Kraus1999,Olea2005}.  Most of these difficulties come from the generality with which the algorithm was developed: it applies to any foliation of AdS, including those with complicated boundary geometries, such that the boundary counterterms contain terms such as $\nabla \R$.
\\\\
A much more straightforward approach to find renormalizing counterterms consists of first choosing a specific foliation, based on the problem of interest, before finding the linear combination of surface curvature scalars which eliminate the divergences \cite{Kraus1999}.  This is how, in practice, the method of counterterms is actually used.  A drawback to this second technique is the limited range of applicability of its results: the counterterms found are no longer valid if the spacetime is foliated differently, and one must perform this tedious task anew for every new problem.
\\\\
We provide a middle-ground solution.  In the spirit of AdS / CFT, one is often interested in boundary geometries which are conformally flat, such as $R^d, S^d, H^d, S^m \times H^{d-m}$ and so on. One may therefore look for a simpler algorithm that can yield counterterms valid for this restricted, yet still fairly large, interesting subset of boundaries.  Our approach will embody the spirit of the second technique described earlier, which we will modify in such a way that the counterterms be valid for every foliation of AdS having one of the boundaries listed above.  Restricting ourselves to those boundaries does not restrict the AdS spacetime itself: rather, it puts restrictions on the kind of conformal field theories our bulk spacetime can be dual to.  These theories live on a cutoff surface which is very close to the boundary, but which may be highly curved and complicated if we so desire; our results are valid when the CFTs live on one of the geometries written above, which are by far the most popular in the literature.  We ultimately provide, as a future reference, expressions for the counterterms up to sixth order in the Riemann tensor.
\\\\
The duality between, on one hand, central charges and couplings in the field theory and, on the other, parameters in the gravitational theory has been explored through the trace anomaly \cite{Skenderis1998,Skenderis1998b}.  In particular, one learns that General Relativity is only dual to those conformal field theories for which all the central charges are equal.  This is simple to understand: GR does not have enough free parameters to account for the ratios between these charges.  In order to probe a larger space of field theories, one must expand to more general theories of gravity which contain more free parameters, such as Lovelock theories.  These additional central charges have recently been investigated holographically  \cite{Sinha2010,Sinha2011} and the compilation of a dictionary has begun \cite{Buchel2009,Myers2010,Camanho:2009vw,Camanho:2009hu,Camanho:2010ru,deBoer:2009pn,deBoer:2009gx}, translating the couplings of this theory into those of the conformal field theory.  Because the Lovelock Lagrangians contain terms of higher order in the Riemann tensor, the algorithm from \cite{Kraus1999} described earlier, while still applicable, becomes yet even more complicated than in the Einstein-Hilbert case, such that generic counterterms for Lovelock theories have yet to be calculated, as pointed out in \cite{Dehghani2006}.  The simpler technique which we will use here will allow us to take on this issue.
\\\\
We begin in Section \ref{sec:method} by describing the general form of the counterterms as well as the technique we will use to derive them.  In Section \ref{sec:gr}, we will go through the derivation of these counterterms for the case of General Relativity, which will exemplify the use of the method, while in Section \ref{sec:ll}, we will apply these ideas to Lovelock theories.  Finally, we will provide a simple application of these counterterms by calculating the entanglement entropy of a spherical surface, following the construction of \cite{Casini2011}, in section \ref{sec:entent}.  In Section \ref{sec:disc}, we will discuss some of the consequences of these counterterms with regards to applications and thermodynamics in general, and comment on their geometric meaning.  Appendix \ref{app:ll} contains some formulas for Lovelock gravity which are difficult or impossible to find in the literature, while in Appendix \ref{app:paddy}, we discuss a curious result whereby one may calculate a renormalized action starting only from a vanishing bulk action and no boundary counterterm.  We comment on renormalizing bulk actions alone (with no boundary terms) in asymptotically AdS spacetimes, and discuss $F(R)$ gravity, in Appendix \ref{app:f(r)} and finally write the expressions for the counterterms in Appendix \ref{ct}.

\section{Form of the counterterms \label{sec:method}}
The setup is as follows: we consider vacuum AdS with scale $L$ and some UV cutoff boundary at a finite radius $r$.  Then, we will look for surface counterterms on this boundary, built only from quantities intrinsic to the boundary, which will cancel out the divergences in the action when the limit $r/L \rightarrow \infty$ is taken.  This is done by expanding the action in powers of $r$ and adding counterterms which cancel the positive powers of $r$ order by order such that, at the end of the procedure, we are left with a finite action.
\\\\
Working in the vacuum of AdS may not seem particularly useful: we are generally interested in spacetimes which are asymptotically locally AdS but which are not AdS vacuum.  However, these same counterterms that we will derive in the vacuum case will also cancel out the divergences that appear in other bulk spacetimes which asymptote to AdS with this same boundary \cite{Kraus1999}, as it is clear that these are all UV divergences with a universal form.  Therefore, our goal is to find counterterms which renormalize gravitational actions in vacuum AdS.
\\\\
Motivated by our goal to study conformal field theories on conformally flat backgrounds, we will restrict ourselves to conformally flat boundaries of the form $S^m \times H^{d-m}$, where $H^{d-m}$ corresponds to the hyperbolic hyperplane; such boundaries were also recently considered in lower-dimensional settings \cite{Jatkar:2011ue}.  More precisely, our bulk metric will take the form
\myeq{ \label{eqn:metric}
ds^2 &= \frac{dr^2}{1 + \frac{r^2}{L^2}} + \left( r^2 + L^2 \right) dH^2_{d-m} + r^2 d\Omega^2_m ~,
}
where $dH^2$ and $d\Omega^2$ are the line elements for the unit-curvature hyperbolic hyperplane and sphere:
\myeq{
dH^2_{d-m} &= du_1^2 + \sinh^2u_1 du_2^2 + \sinh^2u_1 \sin^2u_2 du_3^2 + \sinh^2u_1 \sin^2u_2 \sin^2u_3 du_4^2 + \ldots \\
d\Omega^2_{m} &= d\theta_1^2 + \sin^2\theta_1 d\theta_2^2 + \sin^2\theta_1 \sin^2\theta_2 d\theta_3^2 + \sin^2\theta_1 \sin^2\theta_2 \sin^2\theta_3 d\theta_4^2 + \ldots ~~.
}
The dual CFT lives on the $r/L \rightarrow \infty$ boundary of this spacetime, which is conformally flat since the radii of $S^m$ and $H^{d-m}$ agree in this limit.  However, in order to calculate the counterterms, we need to perform an expansion in $(r/L)^{-1}$; this forces $(r/L)$ to remain a finite quantity throughout the calculations, meaning that the boundary metric will in effect be
\myeq{ \label{eqn:metric2}
ds^2 &= \left( r^2 + L^2 \right) dH^2_{d-m} + r^2 d\Omega^2_m ~.
}
\\\\
Working with this family of foliations $(\ref{eqn:metric})$ will be extremely useful for two reasons.  First, it will provide us with the additional equations we require in order to uniquely set coefficients in the counterterms.  Indeed, these will be of the form
\myeq{ \label{eqn:ct}
I_{ct} = \int d^d x \sqrt{h} \left\{ \alpha_{0,1} + \alpha_{1,1} {\cal R} + \alpha_{2,1} {\cal R}^2 + \alpha_{2,2}{\cal R}^{ab}{\cal R}_{ab} + \cdots \right\}~~;
}
where ${\cal R}_{abcd}$ (in script notation) corresponds to the boundary Riemann tensor and $h_{ab}$ is the induced metric at finite $r$.  Our goal is to find the coefficients $\alpha_{i,j}$, which we do by expanding both this counterterm $(\ref{eqn:ct})$ and the non-regulated bulk and boundary actions in powers of $r$, and matching the coefficients of $r^n$.  Working with a single foliation, it is possible to uniquely set $\alpha_{0,1}$ and $\alpha_{1,1}$.  However, because the counterterm contains two terms of order ${\cal R}^2$, we need at least two independent equations to uniquely set the coefficients $\alpha_{2,1}$ and $\alpha_{2,2}$.  Each foliation (that is, each value of $m$) will provide us with one such equation, up to $m \leftrightarrow (d-m)$.  Working with the generic foliation $H^{d-m} \times S^m$ will therefore provide us with the ever-increasing number of independent equations we will need to uniquely set the higher-order coefficients.
\\\\
There is another important advantage to working with the family of metrics given by $(\ref{eqn:metric})$: by considering a single family of foliations characterized by an integer $m$, we can construct an extremely simple computational framework which can calculate any scalar quantity (built from either the extrinsic curvature or the boundary Riemann tensor) on the boundary, for any $m$.  Indeed, because the boundary is the product of conformally flat Einstein manifolds, the Weyl tensor vanishes on each part and we find, on either $H^{d-m}$ or $S^m$, that the Riemann tensor can be written in terms of the Ricci tensor:
\myeq{ \label{eqn:rabcd}
\R^a_{\phantom{a}bcd} = \frac{1}{p-1} \left( \delta^a_c \R_{bd} - \delta^a_d \R_{bc} \right),
}
where $p$ is the dimension of the submanifold (so either $p=m$ or $p=d-m$).  This will dramatically simplify the calculations since it implies that boundary scalars can be written as linear combinations of terms of the form $\Tr(R^n K^m)$, meaning the trace of the matrix multiplication of $n$ Ricci tensors and $m$ extrinsic curvatures.  Writing $\left\{ \alpha, \beta, \gamma, \delta \right\}$ as $S^m$ indices and $\left\{\mu, \nu, \rho, \sigma \right\}$ as $H^{d-m}$ indices, we can illustrate this idea by calculating
\myeq{
K^{ac} R^{bd} \R_{abcd} &= K^{\alpha \gamma} \R^{\beta \delta} \R_{\alpha \beta \gamma \delta} + K^{\mu \rho} \R^{\nu \sigma} \R_{\mu \nu \rho \sigma}\\
&= \frac{1}{m-1} \left[ \Tr_S(K) \Tr_S(\R^2) - \Tr_S(K \R^2) \right] + \frac{1}{d-m-1} \left[ \Tr_H(K) \Tr_H(\R^2) - \Tr_H(K \R^2) \right]~,
}
where the subscripts $H$ and $S$ mean that we're tracing only over the hyperbolic or spherical indices.
\\\\
The Ricci tensor on each submanifold is given by $R_{ab} = \pm\frac{ (p-1)}{R_0^2} g_{ab}$, where $R_0$ and $p$ are the radius and dimension of the submanifold and where we take the positive sign for the spherical submanifold and the negative for the hyperbolic one.  Therefore, we can easily calculate, with $x \equiv L^2/r^2$:
\myeq{ \label{eqn:partialtrace}
\Tr_S(\R^a K^b) &= \frac{x^a}{L^{2a+b}}\left[ (1+x)^{b/2} m (m-1)^a  \right]\\
\Tr_H(\R^a K^b) &= \frac{x^a}{L^{2a+b}(1+x)^{a+\frac{b}{2}}}\left[ (-1)^a (d-m)(d-m-1)^a \right];
}
the total trace is simply defined as the sum of these two quantities:
\myeq{ \label{eqn:totaltrace}
\Tr(\R^a K^b) = \frac{x^a}{L^{2a+b}(1+x)^{a+\frac{b}{2}}}\left[ (1+x)^{a+b} m (m-1)^a + (-1)^a (d-m)(d-m-1)^a \right] ~.
}
Any contraction of an arbitrary number of boundary Riemann tensors and extrinsic curvatures, on our hypersurface of constant $r$, can be calculated using equations $(\ref{eqn:rabcd})$ and $(\ref{eqn:partialtrace})$, which is quite extraordinary.  This will be crucial when investigating Lovelock theories, whose Gibbons-Hawking-York terms involve incredibly complex contractions (see Appendix \ref{app:ll}) of the extrinsic curvature and the Ricci and Riemann tensors.
\\\\
An important consequence of equation $(\ref{eqn:totaltrace})$ is that the trace of an odd number of Ricci tensors is not independent of the other quantities to leading order in $x$.  For example, we can write
\myeq{
\Tr(\R^3) = \frac{1}{d(d-1)} \left[ (2d-1) \Tr(\R) \Tr(\R^2) - \Tr(\R)^3 \right] ~.
}
Hence, such terms will not appear in our counterterms.  Note also that the counterterms which we will consider will be intrinsic quantities only: these correspond to the $b=0$ case of equation $(\ref{eqn:partialtrace})$.  However, terms with $b \neq 0$ will appear in the surface terms (equivalent to the Gibbons-Hawking-York term for GR) of Lovelock theories.

\section{General Relativity \label{sec:gr}}
Let's begin with a detailed derivation of the counterterms for General Relativity as a preamble to the more complex Lovelock calculations in the following section.  In Anti-deSitter space, the Einstein-Hilbert bulk action is given by
\myeq{
I_{\bulk} = \int d^{d+1} x \sqrt{-g} \left( \frac{d(d-1)}{L^2} + R \right) ~,
}
where the first term corresponds to the cosmological constant and $L$ is the AdS curvature scale.  Note that here and for the rest of the paper, we will drop the prefactor $(16\pi G)^{-1}$ for notational simplicity.
\\\\
The divergences that we are trying to remove come entirely from the asymptotic region of the spacetime \cite{Kraus1999}.  As such, it suffices for us to consider counterterms which will renormalize the action for AdS spacetime itself; these same counterterms will then naturally also renormalize the action for asymptotically AdS spacetimes with an $S^m \times H^{d-m}$ boundary.  Using the fact that AdS is maximally symmetric, meaning $R_{abcd} = \frac{-1}{L^2} \left( g_{ac} g_{bd} - g_{ad} g_{bc} \right)$, and relating the integral over the bulk to the integral over the boundary, we calculate the bulk action
\myeq{ \label{eqn:gr_bulk}
I_{\bulk} &= \frac{-2d}{L^2} \int d^{d+1}x \sqrt{-g}\\
&= \frac{-2d}{L^2}  \sigma  \int dr \left( 1 + \frac{r^2}{L^2} \right)^{-1/2} \left( r^2 + L^2 \right)^{\frac{d-m}{2}} r^m\\
&= \frac{-2d}{L^2} \int d^d x \sqrt{h} \left\{ \frac{1}{r^m \left( 1 + \frac{r^2}{L^2} \right)^{\frac{d-m}{2}}} \int dr  \left( 1 + \frac{r^2}{L^2} \right)^{\frac{d-m-1}{2}}r^m \right\} ~,
}
where $h_{ab}$ corresponds to the induced metric on a hypersurface located at a finite value of $r$ and $\sigma$ is the divergent integral over the unit $m$-sphere and $(d-m)$-hyperbolic hyperplane.  Although writing the bulk action as a boundary integral in the above equation may seem peculiar, it will let us perform all our computations on the integrand alone without ever having to perform the boundary integral over $\int d^d x \sqrt{h}$.
\\\\
For a manifold with boundary such as the one we are considering here, a surface term must be added to the action to ensure that its variational principle is well-defined.  For the Einstein-Hilbert action, this boundary term is known as the Gibbons-Hawking-York term, which is simply $I_{\sur} = 2 \int d^d x \sqrt{h} K$; using equation $(\ref{eqn:totaltrace})$, we can write this term explicitly as
\myeq{ \label{eqn:gr_sur}
I_{\sur} = \int d^d x \sqrt{h} \frac{2}{L \left(1+\frac{L^2}{r^2} \right)^{1/2}} \left[ \left(1+\frac{L^2}{r^2} \right)m + (d-m) \right] ~.
}
The total action is $I = I_{\bulk} + I_{\sur}$ from equations $(\ref{eqn:gr_bulk})$ and $(\ref{eqn:gr_sur})$.  Working order by order (in either $x=L^2/r^2$ or the Ricci tensor, which are of the same order), we look for the coefficients $\alpha_{i,j}$ from equation $(\ref{eqn:ct})$ which ensure that the divergent part of the total action is canceled out by the counterterm for every value of $m$ (that is: every foliation).  To zeroth order, only one constant needs to be found --- $\alpha_{0,1}$ --- such that we only need to consider one foliation, which will be $m=d$ for simplicity.  Setting $I = I_{\ct}$ and truncating to zeroth order in $x$ yields
\myeq{
\underbrace{\frac{2(d-1)}{L}}_{I} = \underbrace{\alpha_{0,1}}_{I_{\ct}} ~.
}
To first order, we once again only need to consider a single foliation, $m=d$, and our single equation is
\myeq{
\underbrace{\frac{d(d-1)}{(d-2)L}}_I = \underbrace{\alpha_{1,1} \frac{d(d-1)}{L^2}}_{I_{\ct}}~,
}
obviously yielding
\myeq{
\alpha_{1,1} = \frac{L}{d-2}~.
}
To second order, we find our first non-trivial system of equations as we have two constants to solve for.  Looking at the two foliations $m=d$ and $m=d-1$ leaves us with two independent equations and two constants to find:
\myeq{
m=d & \rightarrow \frac{-d(d-1)}{4L(d-4)} = \alpha_{2,1} \frac{d^2(d-1)^2}{L^4} + \alpha_{2,2} \frac{d(d-1)^2}{L^4} \\
m=d-1 & \rightarrow \underbrace{\frac{-(d-1)}{4L}}_I = \underbrace{\alpha_{2,1} \frac{(d-1)^2(d-2)^2}{L^4} + \alpha_{2,2} \frac{(d-1)(d-2)^2}{L^4}}_{I_{\ct}}~,
}
yielding the solution
\myeq{
\alpha_{2,1} &= \frac{-d L^3}{(d-4)(d-2)^2} \\
\alpha_{2,2} &= \frac{L^3}{(d-4)(d-2)^2}~.
}
Combining the above calculations and rewriting slightly leads to a counterterm up to order $\R^2$:
\myeq{
I_{\ct} = \int d^d x \sqrt{h} \left( \frac{2(d-1)}{L} + \frac{L}{d-2} {\cal R} + \frac{L^3}{(d-4)(d-2)^2} \left[ \R^{ab} \R_{ab} - \frac{d}{4(d-1)}{\cal R}^2 \right]  + \ldots \right) ~;
}
this has previously been calculated in \cite{Kraus1999} and \cite{Rob1999}.  It is simple to keep going, however, and we can continue this procedure to arbitrary order; we provide the equation to sixth order in $(\ref{eqn:gr_ct})$.  This counterterm can be subtracted from the gravitational action to yield a finite action in asymptotically AdS spacetimes with boundaries given by $S^m \times H^{d-m}$, which are highly interesting choices in light of the AdS / CFT correspondence.  In other words, the following action is finite:
\myeq{
I_{\text{GR,ren}} = \int d^{d+1} x \sqrt{-g} \left( \frac{d(d-1)}{L^2} + R \right) + 2 \int d^d x \sqrt{h} K - I_{\ct} ~.
}
Because of the powers of $r$ which are hidden inside $\sqrt{h}$, the counterterm $(\ref{eqn:gr_ct})$ is really an expansion in $r$, with $\sqrt{h}\R^i$ being of order $r^{d-2i}$.  Therefore, the series should be truncated at order $n$, where $d=2n+1,2n+2$ \cite{Rob1999,Kraus1999}, to ensure that only the divergences get subtracted.  For even $d=2m$, this means that a finite term will be left over in the action; indeed, this term cannot be subtracted since its corresponding counterterm would have to be proportional to $\sqrt{h}\R^m$ which, as can be seen in equation $(\ref{eqn:gr_ct})$, is proportional to $\frac{1}{d-2m}$ and would therefore diverge.  This finite term is the gravitational anomaly \cite{Kraus1999}.
\\\\
While the previously known results in the literature, which contained terms up to third order in the Riemann tensor, allowed one to study up to eight bulk dimensions, the counterterm we provide works in up to fourteen dimensions, and can easily be expanded to work in an arbitrary number of dimensions.  This therefore provides much greater freedom in studying higher-dimensional field theories.

\section{Lovelock Gravity \label{sec:ll}}
A common application for the counterterm method is holography: the thermodynamics of asymptotically AdS spacetimes are dual to the thermodynamics of some field theory living on this boundary.  However, this duality is greatly constrained by the fact that General Relativity is dual only to those conformal field theories where all the central charges are equal; this is essentially caused by the lack of free parameters.  Although this is not a problem when working with two-dimensional conformal field theories, whose universality class is uniquely defined by a single central charge $c$, it becomes an important limiting factor when trying to study higher-dimensional field theories which may have multiple central charges.  A more general theory of gravity, built from Lovelock Lagrangians \cite{Lovelock:1971yv} (for a review of such Lagrangians, see Ch. 15 of \cite{Paddybook}), can help with this issue as it provides additional parameters which can be dual to CFT couplings.  In particular, the coupling in front of the Gauss-Bonnet term $L_2$, described below, has recently been precisely linked to the ratio between the central charges of the dual higher-dimensional CFT \cite{Buchel2009,Myers2010}.  Although briefly studied in the case of four-dimensional Gauss-Bonnet gravity \cite{Cvetic:2001bk,Astefanesei:2008wz}, these counterterms are not known for Lovelock theories in $d$ dimension\cite{Dehghani2006}; therefore, we will follow the same steps as in the previous section to fill this gap in the literature.
\\\\
These Lovelock Lagrangians are the natural generalization of the Einstein-Hilbert Lagrangian to higher dimensions.  Indeed, just as the Einstein-Hilbert Lagrangian is the Euler density of a two-dimensional manifold, the $m$th Lovelock Lagrangian $L_m$ is the Euler density of a $2m$-dimensional manifold.  Thus, the quantity $L_m$ vanishes in less than $2m$ dimensions, and is a total derivative in exactly $2m$ dimensions \cite{YalePaddy} since it is topological; in particular, this means that the only Lovelock Lagrangian which is not a total derivative in $4$ dimensions is the $m=1$ Einstein-Hilbert term.  Moreover, the $m$-th Lovelock Lagrangian is the unique scalar of order $m$ in the Riemann tensor whose equations of motion are second order: they are specifically built so that the fourth-order terms cancel out.  These Lagrangians are given by
\myeq{ \label{eqn:lm}
L &= \sum_m \lambda_m L_m \\
L_m &= \delta^{A_1 B_1 A_2 B_2 \cdots A_m B_m}_{C_1 D_1 C_2 D_2 \cdots C_m D_m} R_{A_1 B_1}^{C_1 D_1} R_{A_2 B_2}^{C_2 D_2} \cdots R_{A_m B_m}^{C_m D_m}~,
}
where
\myeq{
\delta^{A_1 B_1 A_2 B_2 \cdots A_m B_m}_{C_1 D_1 C_2 D_2 \cdots C_m D_m} = \delta^{[A_1}_{C_1} \delta^{B_1}_{D_1} \cdots \delta^{B_m]}_{D_m}
}
is the totally antisymmetric determinant tensor.  In particular, the $m=1$ cases leads to the Einstein-Hilbert Lagrangian $L_1 = R$, while $L_2 = R^2 - 4R_{ab}R^{ab} + R_{abcd}R^{abcd}$ is the Gauss-Bonnet Lagrangian.  Examples of Lovelock Lagrangians for larger values of $m$ can be found in Appendix \ref{app:ll}.
\\\\
Thanks to the simplicity of our method for finding counterterms, we are able to calculate them for any Lovelock theory, and to arbitrary order.  The first step is calculating the bulk action
\myeq{ \label{action_ll}
I_{\bulk ,m} = \int d^{d+1} x \sqrt{-g} \left( L_m - 2 \Lambda_m \right) ~.
}
As in the previous section, the fact that the bulk spacetime is maximally symmetric makes evaluating $L_m$ a trivial matter.  However, because the equations of motion from $(\ref{action_ll})$ are now different than in the case of General Relativity, we need to find the cosmological constant $\Lambda_m$ which ensures that the metric $(\ref{eqn:metric})$ is a solution.  While varying the Lovelock Lagrangian in equation $(\ref{action_ll})$ appears at first to be a daunting challenge, one finds a surprisingly simple result \cite{PaddyLL}.  In vacuum, the equations of motion can be written as
\myeq{
0 &= P_a^{\phantom{a}bcd}R_{ebcd} - \frac{1}{2}g_{ae} L_m + \Lambda g_{ae} \\
P_{cd}^{\phantom{cd}ab} &= m \delta^{a b a_2 b_2 a_3 b_3 \cdots a_m b_m}_{c d c_2 d_2 c_3 d_3 \cdots c_m d_m} R^{c_2 d_2}_{a_2 b_2} \cdots R^{c_m d_m}_{a_m b_m}~.
}
After some algebra, one finds that this leads to
\myeq{ \label{eqn:ll_bulk}
L_\bulk = L_m - 2\Lambda = \frac{2m(-1)^m}{L^{2m}}d(d-1)(d-2)\cdots(d-2m+2);
}
In particular, we recover the result from the previous section, the prefactor in equation $(\ref{eqn:gr_bulk})$, for $m=1$: $L_1 - 2\Lambda = \frac{-2d}{L^2}$.  The bulk action is then simply given by $I_\bulk = L_\bulk \int d^{d+1} x \sqrt{-g}$.
\\\\
Just as a Gibbons-Hawking surface term must be added to the Einstein-Hilbert action to make the variational principle well-defined, a surface countribution must be added to our Lovelock Lagrangian.  The topological nature of these Lagrangians will help us find this term: in particular, the Lagrangians are closed and locally exact in $d=2m$ dimensions.  The Chern-Simons form $Q$ is defined to embody this quality: given two connections $\Gamma_1$ and $\Gamma_2$, we have dimensions $dQ(\Gamma_1,\Gamma_2) = L_m(\Gamma_1)-L_m(\Gamma_2)$.  Then, index theorems \cite{Eguchi, Nakahara} tell us that for a manifold with boundary, the Euler characteristic $\chi$ must be supplemented by a boundary contribution: $\chi = \int (L_m(\Gamma) + dQ(\Gamma_0,\Gamma) )$; pictorially, the $dQ$ in this equation replaces the bulk Lagrangian $L_m(\Gamma)$ with $L_m(\Gamma_0)$ where $\Gamma_0$ is the connection for a product metric having the same boundary.  In common language, this Chern-Simons term $Q$ can be written \cite{Myers1987, Olea2005}
\myeq{ \label{eqn:qm}
Q_m = 2m \int_0^1 dt \delta^{a_1 a_2 \cdots a_{2m-1}}_{b_1 b_2 \cdots b_{2m-1}} K^{b_1}_{a_1} \left( \frac{1}{2} R^{b_2 b_3}_{a_2 a_3} - t^2 K^{b_2}_{a_2} K^{b_3}_{a_3} \right) \cdots \left( \frac{1}{2} R^{b_{2m-2} b_{2m-1}}_{a_{2m-2} a_{2m-1}} - t^2 K^{b_{2m-2}}_{a_{2m-2}} K^{b_{2m-1}}_{a_{2m-1}} \right);
}
the Einstein-Hilbert and Gauss-Bonnet surface terms are thus:
\myeq{
Q_1 &= 2 \Tr(K) \\
Q_2 &= 4\Tr(\R)\Tr(K) - 8 \Tr(K \R) - \frac{4}{3} \Tr(K)^3 + 4\Tr(K)\Tr(K^2)  - \frac{8}{3} \Tr(K^3),
}
while larger surface terms can be found in Appendix \ref{app:ll}.  The Gibbons-Hawking-like surface term will be the dimensionally-extended version of $Q_m$ and will precisely cancel out normal derivatives of the metric on the boundary, thus ensuring that the variational principle is well defined.  This has long been known for the $m=1$ Einstein-Hilbert case, and has also been done in detail for the $m=2$ Gauss-Bonnet term \cite{Bunch}.  These boundary terms have in particular been used to find the Israel conditions \cite{Davis} and Friedmann equations \cite{Gravanis} for Einstein-Gauss-Bonnet gravity.  For a more in-depth discussion, we refer the reader to \cite{YalePaddy} and references therein.
\\\\
Upon using equation $(\ref{eqn:totaltrace})$ to write these $Q_m$ explicitly as functions of $x = \frac{L^2}{r^2}$, the total action is found by integrating the bulk action from equation $(\ref{eqn:ll_bulk})$, as was done for General Relativity in equation $(\ref{eqn:gr_bulk})$, and adding the boundary integral of $Q_m$ from equation $(\ref{eqn:qm})$.  Proceeding in precisely the same way as in Section \ref{sec:gr}, we solve linear equations to find the surface counterterm that we should add in order to make the action finite.  As in the previous section, we find the counterterms up to sixth order for future reference: they are given in equations $(\ref{eqn:ll_ct})$ and $(\ref{eqn:ll_ct})$ for second-order and third-order Lovelock gravity, respectively.
\\\\
In practice, one should combine many of the above counterterms.  For example, in order to regularize an action with Einstein-Hilbert, Gauss-Bonnet and $m=3$ Lovelock terms,
\myeq{
I = \int \left[ \lambda_1(R - 2 \Lambda_1) + \lambda_2(L_2 - 2\Lambda_2) + \lambda_3(L_3 - 2\Lambda_3) \right] + \oint \left[ \lambda_1(2K) + \lambda_2(Q_2) +\lambda_3(Q_3) \right] ~,
}
one must combine the counterterms from equation $(\ref{eqn:gr_ct})$ with those from equations $(\ref{eqn:ll_ct})$ and $(\ref{eqn:ll_ct2})$.  These equations are the main results of this paper.  As discussed at the end of Section \ref{sec:gr}, they can renormalize the action in up to fourteen boundary dimensions.

\section{Simple Application: Entanglement Entropy \label{sec:entent}}
Let us now demonstrate a practical implementation of these results to holographic calculations of entanglement entropy \cite{Ryu1,Ryu2,Ryu3}.  It has recently been shown \cite{Casini2011} that the entanglement entropy for a spherical surface in a CFT is equal to the Wald entropy of the ``topological black hole'' consisting of the hyperbolic foliation of AdS$_{4+1}$:
\myeq{ \label{newmetric}
ds^2 = -\left( \frac{r^2}{L^2}-1\right)\frac{L^2}{R^2} dt^2 + \frac{dr^2}{\frac{r^2}{L^2}-1} + r^2 dH_3^2 \,.
}
We will not go into the details of this construction, which can be found in \cite{Casini2011}. Rather, our aim is to provide a simple example where the counterterms may be used.  If we take our gravity action to be Einstein-Gauss-Bonnet:
\myeq{
I = \frac{1}{2 l_P^3} \int d^5 x \sqrt{-g} \left[ -\frac{12}{L^2 f_\infty} + R + \frac{\lambda f_\infty L^2}{2} L_2 \right] + \frac{1}{2l_P^3} \int d^4 x \sqrt{h} \left[ 2 K + \frac{\lambda f_\infty L^2}{2} Q_2 \right]\,,
}
where $f_\infty$ obeys $1 - f_\infty + \lambda f_\infty^2 = 0$, then the entanglement entropy will be proportional to the central charge $a$ of the CFT \cite{Casini2011} which, using the standard AdS/CFT dictionary \cite{Buchel2009,Myers2010}, can be written
\myeq{ \label{a}
a = \pi^2 \left( \frac{L}{l_P}\right)^3 \left( 1 - 6 \lambda f_\infty \right) \,.
}
Using the counterterms calculated in the previous sections, we can present an alternate and elegant derivation of this result.  Instead of directly computing the Wald entropy of the horizon, as is done in \cite{Casini2011}, we will compute the free energy ${\cal F}$ of the spacetime by integrating the action.  In particular, this will reproduce the entropy result through $S \propto \partial {\cal F} / \partial T$.
\\\\
This free energy will necessarily diverge due to the large-$r$ bound of integration, and the standard technique to deal with this --- background subtraction --- is ambiguous since our spacetime is already (a particular coordinate choice of) vacuum AdS.  The introduction of counterterms allows us to deal with the large-$r$ divergences and calculate the free energy:
\myeq{ \label{free}
\beta {\cal F} &= \frac{1}{2 l_P^3} \int d^5x \sqrt{-g} \left[ -\frac{12}{L^2 f_\infty}+ R + \frac{\lambda f_\infty L^2}{2} L_2 \right] + \frac{1}{2l_P^3} \int d^4 x \sqrt{h} \left[ 2 K + \frac{\lambda f_\infty L^2}{2} Q_2 \right] \\
&~~~~- \frac{1}{2l_P^3} \int d^4 x \sqrt{h} \left[ \frac{6}{L} + \frac{L}{2} {\cal R} + \frac{\lambda f_\infty L^2}{2} \left( \frac{-8}{L^3} + \frac{2}{L} {\cal R} \right) \right] \\
&= \frac{-\left(1-6 \lambda f_\infty \right) L^3}{l_P^3R} \beta H \,,
}
where $H = \int dH_3$ is the area of the hyperbolic slice.  The second line in this expression corresponds to the counterterms from equations $(\ref{eqn:gr_ct})$ and $(\ref{eqn:ll_ct})$.  These are truncated to order ${\cal R}$ because we are working with $d=4$ for simplicity, and the constant term which they introduce was thrown out.  Since the temperature $T = \beta^{-1}$ for the line element $(\ref{newmetric})$ is given by \cite{YM}
\myeq{
T &= \frac{L}{4\pi R} \partial_r \left( \frac{r^2}{L^2}-1\right)\Big|_{r=L} = \frac{1}{2\pi R} \,,
}
we can calculate the entropy density:
\myeq{
s = S/H = -\frac{\partial \left( {\cal F}/H\right)}{\partial T} = \frac{2}{\pi} a \,,
}
where we used equation $(\ref{a})$.  This result is in agreement with \cite{Casini2011}.

\section{Discussion \label{sec:disc}}
A popular motivation for studying renormalization counterterms in asymptotically AdS spacetimes is to understand the thermodynamics of their associated conformal field theories.  Indeed, the action of a bulk theory of gravity evaluated on an asymptotically AdS background provides us, through the usual ideas of black hole thermodynamics, with the free energy of this spacetime.  The AdS / CFT correspondence then links this free energy with that of a thermal vacuum of the dual field theory living on the boundary.  Although this free energy --- and its associated entropy --- is generally divergent, the method of counterterms not only provides a consistent and mostly unambiguous way of calculating the renormalized bulk free energy, it does so in a way which is analogous to the counterterm renormalization technique used on the CFT side.  Because our technique is easily applicable to any gravitational Lagrangian, as demonstrated by our application to Lovelock Lagrangians, a large number of conformal field theories can be studied in this manner.  Indeed, previous work on counterterms has essentially focused on General Relativity and on the first two or three terms in the counterterm expansion; this not only limits the number of dimensions that the dual field theory may have, but also constrains us to look only at a small section of the space of field theories, where all central charges are equal.  By generalizing to both higher-order counterterms and wider gravitational theories, we have lifted both of these constraints, at the very small cost of requiring the field theories to live on conformally flat Einstein spacetimes.
\\\\
An important lesson, which was also pointed out in \cite{Kraus1999}, is that the renormalized action will see no contribution from surface terms at infinity, such as the Gibbons-Hawking-York surface term, because they are entirely canceled out by the renormalization procedure.   This means that all the information contained in the renormalized action appears to be included in the bulk term alone.  As we exemplify in Appendix \ref{app:paddy}, this can even be true when the bulk term vanishes!
\\\\
In fact, the counterterm plays two roles: it cancels the boundary term at infinity and regularizes the volume integral.  This can be seen by writing the bulk Lagrangian, in a stationary spherically symmetric background, as a total derivative, $L(r) = \partial_r f(r)$, such that the action is:
\myeq{
I &= \int_{r_0}^r L dr + \oint Q\\
&= f(r) - f(r_0) + \oint Q~.
}
Then, up to a constant, the only contribution which will survive the renormalization process is $f(r_0)$: that is, the entire non-trivial contribution to the free energy will come from the horizon.  This is caused by the divergent nature of $f(r)$ and $\oint Q$, both of which being evaluated at infinity, and the counterterm's role will be to cancel them out up to some constant.
\\\\
This means that the boundary terms of the non-renormalized theory are not required in order to find the renormalized action.  In that spirit, we will provide an expression for counterterms which regularize bulk Lagrangians in asymptotically AdS spacetimes in Appendix \ref{app:f(r)}, allowing one to renormalize theories of gravity for which surface terms are not usually used, such as $F(R)$ gravity.  Moreover, this point of view is particularly insightful in that it can shed some light on some alternate counterterms which have been proposed in an effort to understand the geometric meaning of counterterms.  In particular, it has been noticed \cite{Olea2005} that in $2n$ dimensions, another possible counterterm for either Einstein-Hilbert or Lovelock gravity (without Gibbons-Hawking surface terms) is the Chern-Simons form $Q_n$, with a proportionality constant depending on the theory which we are investigating.  One should note, however, that unlike the counterterms we present here, $Q_n$ depends on quantities extrinsic to the boundary. In four-dimensional General Relativity, the following action is finite for some value of $\alpha$:
\myeq{ \label{eqn:Olea1}
I_{ren} = \int \left(R + \frac{6}{L^2} \right) + \alpha \oint Q_2~.
}
It seems very surprising that the surface term for $m=2$ Lovelock gravity, $Q_2$, somehow knows enough about the Einstein-Hilbert action to cancel its divergences.  However, it turns out that this is merely a regularization of the volume integral.  Indeed, because the Euler number in four dimensions is $\chi = \int L_2 + \oint Q_2$, then the regularizing term $\oint Q_2$ in equation $(\ref{eqn:Olea1})$ is merely, up to a constant ($\chi$), the volume integral of the Gauss-Bonnet Lagrangian $L_2$.  However, since both $R=\frac{-12}{L^2}$ and $L_2=\frac{24}{L^2}$ are constants in AdS, then the above action $(\ref{eqn:Olea1})$ can be written $I_{ren} = \int \left( \frac{-6}{L^2} - \alpha \frac{24}{L^4} \right) + \alpha \chi$, which diverges unless $\alpha=\frac{-L^2}{4}$; this value of $\alpha$ is also found in \cite{Olea2005}.  Thus, $Q_2$ regularizes the action in pure AdS.  No issues will arise when we extend the discussion to asymptotically AdS spacetimes with, say, a black hole in the center, since we've seen that all the divergences come from the boundary.  Through a Fefferman-Graham expansion, \cite{Olea2007, Olea2009} shows that this is indeed the case.
\\\\
Because the counterterms are essentially a series in $\sum_i r^{d-2i}$, one may find, depending on whether the dimensionality is even or odd, a finite counterterm.  This introduces an inherent ambiguity as to whether this finite term should be added, as initially pointed out in \cite{Kraus1999,Kraus1999b}.  This problem is akin to the fact that one may add a constant term to the action without affecting the equations of motion.  However, adding this constant term would affect both the free energy and energy density, emphasizing once more that these are only useful as relative quantities.
\\\\
As we showed in section \ref{sec:entent}, these counterterms let us calculate the renormalized free energy without having to worry about the usual ambiguity of background subtraction.  This allowed us to calculate the entropy density for a hyperbolic black hole with inverse temperature $\beta = 2 \pi R$.  This quantity has previously been shown to be equal to the entanglement entropy for a spherical entangling surface \cite{Casini2011} and is proportional to the A-type central charge of the boundary theory.
\\\\
The counterterms that we find in equations $(\ref{eqn:gr_ct})$, $(\ref{eqn:ll_ct})$ and $(\ref{eqn:ll_ct2})$ apply to the boundaries of asymptotically conformally flat AdS spaces which are among the most interesting in the context of AdS / CFT.  Counterterms were previously known to third order in General Relativity, which only renormalize the action in up to $8$ dimensions; by including the counterterms to sixth order, the action remains finite in up to $14$ dimensions.  Moreover, the Lovelock theories of gravity which we considered are dual to richer families of conformal field theories than pure General Relativity.

\acknowledgments{The author would like to thank Rob Myers for many insightful comments and suggestions, as well as Janet Hung, Nima Doroud and Ross Diener for insightful discussions.  Some of the calculations were done with the help of the tensor manipulation software Cadabra \cite{Cadabra1, Cadabra2}. The author is supported by the National Science and Engineering Research Council of Canada.  Research at Perimeter Institute is supported by the Government of Canada through Industry Canada and by the Province of Ontario through the Ministry of Research \& Innovation.
}

\appendix
\section{Lovelock Theory \label{app:ll}}
The Lovelock Lagrangian $L_m$ consists of the Euler density for a compact $2m$-dimensional manifold as well as its corresponding Chern-Simons surface term $Q_m$ \cite{Myers1987, Olea2005} which yields the boundary contribution to the Euler number for a manifold with boundary.  Although the action can be succinctly written as $S = \int L_m + \oint Q_m$ with $L_m$ and $Q_m$ given by equations $(\ref{eqn:lm})$ and $(\ref{eqn:qm})$, these definitions involve the totally antisymmetric tensor $\delta^{a_1 a_2 \ldots a_{2m}}_{b_1 b_2 \ldots b_{2m}}$ which we must expand in order to perform actual computations.  As this can be a tedious exercise in practice, especially for the surface term, the expressions for Lovelock Lagrangians past the Gauss-Bonnet term ($m=2$) are extremely difficult to find in the literature.  We therefore provide some of these here for convenience.  The bulk terms are
\myeq{
L_1 &= R\\
L_2 &= R^2 - 4 R_{ab} R^{ab}\\
L_3 &= R^3 - 12 R R^{ab}R_{ab} + 3 R R^{abcd}R_{abcd} + 16R^{ab}R_a^{\phantom{a} c} R_{bc} + 24R^{ac}R^{bd}R_{abcd} - 24R^{ab}R_a^{\phantom{a} cde}R_{bcde} \\
&~~~+ 2R^{abcd}R_{ab}R^{ef} R_{cdef} - 8R^{abcd} R_{a \phantom{e} c}^{\phantom{a} e \phantom{c} f} R_{bfde}
}
while the surface terms are
\myeq{
Q_1 &= 2K\\
Q_2 &= 4K \R - 8K^{mn}\R_{mn} - \frac{4}{3} K^3 + 4K K^{mn}K_{mn} - \frac{8}{3}K^{mn}K_m^p K_{np}\\
Q_3 &= 6 K \R^2 - 24K \R^{mn}\R_{mn} + 6K\R^{mnpq}\R_{mnpq} - 24K^{mn}\R \R_{mn} + 48K^{mn}\R_m^p\R_{np} + 48K^{mn}\R^{pq}\R_{mnpq} \\
&~~~- 24K^{mn}\R_m^{pqr}\R_{npqr} - 4\R K^3 + 24 K^2 K^{mn}\R_{mn} + 12KK^{mn}K_{mn}\R - 48KK^{mn}K_m^p\R_{np}  \\
&~~~- 24KK^{mn}K^{pq}\R_{mpnq} + 48K^{mn}K_m^pK_n^q\R_{pq} + 48K^{mn}K_m^pK^{qr}\R_{nqpr} + \frac{6}{5}K^5 - 12K^{mn}K_{mn}K^3  \\
&~~~+ 24K^{mn}K_m^pK_{np}K^2 + 18KK^{mn}K_{mn}K^{pq}K_{pq} - 36KK^{mn}K_m^pK_n^qK_{pq}  \\
&~~~- 24K^{mn}K_{mn}K^{pq}K_p^rK_{qr} + \frac{144}{5}K^{mn}K_m^pK_n^qK_p^rK_{qr} ~.
}

\section{A curiosity: renormalized action from a vanishing bulk action \label{app:paddy}}
As discussed in the last paragraph of Section \ref{sec:disc}, a somewhat surprising consequence of these renormalization techniques is that the free energy appears to be entirely determined by the bulk Lagrangian.  Although formulated differently, this was noticed by Padmanabhan \cite{Paddy2002} while studying the thermodynamics of black holes.  Because it shines some light on our discussion, we shall recap some of his arguments.  Consider a stationary spherically symmetric black hole whose line element is given by
\myeq{
ds^2 = -f(r) dt^2 + \frac{dr^2}{f(r)} + dX^2~,
}
where $X$ corresponds to the transverse coordinates which we will drop for the remainder of this discussion.  Then, the Ricci scalar can be written as a total derivative:
\myeq{
R = \nabla_r^2 f(r) - \frac{2}{r^2} \partial_r \left( r(1-f) \right).
}
The (Euclidean) action is now integrated over the exterior region of the spacetime, from the location of the horizon $r_0$ up to some large radius $r$:
\myeq{ \label{eqn:paddyaction}
I &= \frac{-\beta}{4} \int_{r_0}^r dr \left[ -\partial_r (r^2 f') + 2\partial_r (r(1-f)) \right] \\
&= \underbrace{\frac{-\beta}{4} \left[ r_0^2 f'(r_0) - 2r_0 \right]}_{I_{ren}} + \underbrace{\frac{\beta}{4} \left[ r^2 f'(r) - 2r \right]}_{F(r)} ~,
}
where $F(r)$ is a function of the large radius $r$ and will be renormalized away.  Although we did not explicitly add a surface term to our action, one can imagine that $F(r)$ could account for it.  If we denote the period of the Euclidean time by $\beta \equiv \frac{4\pi}{f'(r_0)}$, then the above expression relates the Euclidean action to the free energy through $I_{ren} = S - \beta E$ where $S = \frac{A}{4}$ is the entropy and $E=\sqrt{\frac{A}{16\pi}}$ is the energy.  While interesting, this is all now quite well understood.
\\\\
An important observation, however, is rarely stressed.  Consider now the Schwarzschild black hole, which corresponds to $f(r) = 1 - \frac{2M}{r}$.  Because the Ricci scalar vanishes, the typical calculation for the renormalized Euclidean action is very different than the one given above: we write $I = 2 \oint K$ over the boundary, and subtract the flat-space contribution $I_0 = 2 \oint K_0$ to find the finite action $I - I_0 = 4\pi M^2$.  On the other hand, if we tried to use the above, more general, technique, we would find that the Ricci scalar vanishes because it is the total derivative of a constant...  Yet, following through with the outlined steps, we would recover once more the correct finite action $I_{ren}=4\pi M^2$.  Thus, we are led to a curious result: \emph{even when the bulk action vanishes, it can still contain all the information of the renormalized action}.
\\\\
An astute eye will notice an ambiguity: because the Ricci scalar is the total derivative of a constant, then we can choose that constant to be anything, and not necessarily this $4 \pi M^2$.  This ambiguity actually lets us modify the entropy as we can add a term that depends on $M$ to the term in square brackets in $(\ref{eqn:paddyaction})$ (provided we also subtract that same term from $F(r)$, but that gets thrown out according to the procedure); this is obviously nonsense.  The counterterm technique provides an elegant solution to this puzzle by properly defining what is meant by ``throwing away'' the function $F(r)$ in equation $(\ref{eqn:paddyaction})$ (that is: how much of $F(r)$ comes from the boundary at infinity, which should be removed, and how much of $F(r)$ comes from a constant that we've added, which should be kept so that the entropy remains unchanged).  With this strategy, equation $(\ref{eqn:paddyaction})$ could actually be written
\myeq{
I &= \frac{-\beta}{4} \left[ r_0^2 f'(r_0) - 2r_0 + C \right] + \underbrace{\frac{\beta}{4} \left[ r^2 f'(r) - 2r + C \right] + 2 \oint K}_{F(r)} - \underbrace{\left\{ \frac{\beta}{4} \left[ r^2 f'(r) - 2r\right] + 2 \oint K  \right\}}_{I_{\ct}} \\
&= \frac{-\beta}{4} \left[ r_0^2 f'(r_0) - 2r_0 \right]~,
}
where the arbitrary constant $C$ no longer plagues the final result. By the underbrace denoting $I_{\ct}$, we really mean that $I_{\ct}$ is an expansion in boundary quantities which is equal to the term above the underbrace; the actual form of $I_{\ct}$ would need to be derived through the non-vanishing off-shell bulk action (which will not be vanishing).  The counterterms might therefore provide a procedure for extracting the renormalized action from the bulk action even when it vanishes, and without having to introduce boundary terms.

\section{Regularizing the bulk action alone \label{app:f(r)}}
In light of our discussion and the previous appendix, it is clear that the ideas presented in this paper can be applied to gravitational theories for which we do not generally know the Gibbons-Hawking-York equivalent boundary term.  Indeed, as previously noted, the contribution from this boundary term to the action would have been entirely canceled out by our counterterm.  One example of gravitational theories where surface terms are not generally used are $F(R)$ theories of gravity (see \cite{Sotiriou2010,Nojiri:2010wj} for modern reviews), where the bulk action is some function of the Ricci scalar:
\myeq{
I = \int d^{d+1} x \sqrt{-g} F(R)~.
}
Although these theories are plagued by a number of difficulties, such as an ill-defined initial-value problem, they have received a considerable amount of interest in the literature, particularly in cosmology where one goal is to find a function $F(R)$ which could replace dark energy \cite{Hu2007,Starobinsky2007}; moreover, they are equivalent to scalar-tensor theories of gravity \cite{Paddybook}.  Recently, some of the attention has shifted to studying $F(R)$ gravity in terms of the gauge / gravity duality, where the aim \cite{Liu2010,Caravelli2010} is to understand which holographic features of gravity (such as calculation of various types of entropies, the existence of Cardy's $c$-theorem, etc) are specific to Einstein-Hilbert or Lovelock Lagrangians, and which are more universal.  We therefore provide a simple expression for the counterterm, to fourth order, valid for the family of bulk metrics defined in equation $(\ref{eqn:metric})$:
\myeq{
I_{\ct} &= F \left( \frac{-d(d+1)}{L^2} \right) \int d^d x \sqrt{h} \Bigg(
\frac{L}{d}
-\frac{L^3}{2(d-2)(d-1)d} \\
&+\frac{L^5}{2(d-4)(d-2)^2(d-1)d} \bigg[ 3 d \frac{\R^2}{4(d-1)} - 3 \R^{ab} \R_{ab} \bigg] \\
&+\frac{L^7}{4(d-6)(d-4)(d-2)^3(d-1)^2d^2} \bigg[ \big( -160+160 d-10 d^2-5 d^3 \big) \frac{\R^3}{4(d-1)} + \big(40-70 d+15 d^2\big)\R \R^{ab}\R_{ab} \bigg] \\
&+\frac{L^9}{8(d-8)(d-6)(d-4)(d-2)^4(d-1)^2d} \bigg[ \\
&~~~~\big(20480-48384 d+28672 d^2-5528 d^3+210 d^4+35 d^5\big) \frac{\R^4}{16(d-1)^2 d^2}\\
&~~~~+ \big(2560-8608 d+11352 d^2-5798 d^3+1229 d^4-105 d^5\big) \frac{\R^2 \R^{ab}\R_{ab}}{2(d-1)^2 d^2} \\
&~~~~+\big( -440+114 d+11 d^2 \big) \R^a_b \R^b_c \R^c_d \R^d_a \bigg]
+ \ldots \Bigg)
}
Setting $F(x)=1$, this provides a regularization counterterm for the volume of asymptotically AdS spacetimes.

\section{Expressions for the counterterms} \label{ct}
The counterterm for General Relativity, derived in Section \ref{sec:gr}, is:
\myeq{ \label{eqn:gr_ct}
I_{\ct} &= \int d^d x \sqrt{h} \Bigg( \frac{2(d-1)}{L} + \frac{L}{(d-2)}\R + \frac{L^3}{(d-4)(d-2)^2} \left( {\cal R}^{ab} \R_{ab} - \frac{d}{4(d-1)}{\cal R}^2  \right) \\
& + \frac{L^5}{2 (d-6) (d-2)^3(d-1)^2 d} \bigg[ \frac{1}{4} \left( d^2 + 6d - 8 \right) \R^3 - (d-1)(3d-2)\R \R^a_b \R^b_a \bigg] \\
&+ \frac{L^7}{4(d-8)(d-6)(d-4)(d-2)^4(d-1)} \bigg[ \\
&~~~~\left( -4096+9984 d-5632 d^2+936 d^3-30 d^4-5 d^5 \right)  \frac{\R^4}{16 d^2 (d-1)^2}\\
&~~~~+ \left( -512+1760 d-2344 d^2+1130 d^3-203 d^4+15 d^5 \right) \frac{\R^2 \R^a_b \R^b_a}{2d^2 (d-1)^2} \\
&~~~~+ \left( 136-62 d+3 d^2 \right) \R^a_b \R^b_c \R^c_d \R^d_a \bigg] \\
&+\frac{L^9}{8(d-10)(d-8)(d-6)(d-4)(d-2)^5(d-1)^3d} \bigg[ \\
&~~~~\Big( -163840+903168 d-1734656 d^2+1349248 d^3-463984 d^4+70380 d^5\\
&~~~~~~~~-4752 d^6+77 d^7+7 d^8 \Big) \frac{\R^5}{16(d-1)^2 d^2}  \\
&~~~~+ \Big(-20480+133376 d-344832 d^2+448128 d^3-288428 d^4+90032 d^5\\
&~~~~~~~~-13185 d^6+968 d^7-35 d^8 \Big) \frac{ \R^3 \R^{ab} \R_{ab}}{2(d-1)^2d^2} \\
&~~~~+ \Big(7552-30480 d+28668 d^2-9576 d^3+1097 d^4-33 d^5 \Big) \R \R^a_b \R^b_c \R^c_d \R^d_a \bigg] \\
&+\frac{L^{11}}{8(d-12)(d-10)(d-8)(d-6)(d-4)(d-2)^6(d-1)^3} \bigg[ \\
&~~~~\Big( -15728640+158007296 d-598376448 d^2+1080991744 d^3-998355456 d^4+493757056 d^5 \\
&~~~~~~~~-131566512 d^6+18413972 d^7-1333692 d^8+45543 d^9-378 d^{10}-21 d^{11} \Big) \frac{1}{64(d-1)^4 d^4} \R^6 \\
&~~~~+\Big( -3932160+43433984 d-194895872 d^2+458334208 d^3-615241984 d^4+476404592 d^5\\
&~~~~~~~~-211709588 d^6+52722304 d^7-7059991 d^8+498441 d^9-17865 d^{10}+315 d^{11}\Big) \frac{\R^4 \R^{ab} \R_{ab}}{16(d-1)^4d^4}  \\
&~~~~+ \Big( 1449984-9623040 d+24449280 d^2-26776016 d^3+14306460 d^4-3857428 d^5\\
&~~~~~~~~+507861 d^6-30858 d^7+645 d^8  \Big) \frac{1}{4(d-1)^2 d^2} \R^2 \R^a_b \R^b_c \R^c_d \R^d_a \\
&~~~~+ \Big( -157056+261136 d-156092 d^2+41432 d^3-4945 d^4+285 d^5-6 d^6\Big) \R^a_b \R^b_c \R^c_d \R^d_e \R^e_f \R^f_a
\bigg] + \ldots \Bigg)~.
}

The counterterms for $m=2$ and $m=3$ Lovelock theories, derived in Section \ref{sec:ll} are given by
\myeq{ \label{eqn:ll_ct}
I_{m=2,\ct} &= \int d^d x \sqrt{h} \Bigg( -\frac{4 (d-3) (d-2) (d-1)}{3 L^3} + \frac{2(d-3)}{L} \R \\
&+ \frac{L}{(d-2)(d-4)} \bigg[ 3d(d-3) \frac{\R^2}{2(d-1)}
-6(d-3)\R^{ab}\R_{ab} \bigg] \\
&+\frac{L^3}{3(d-6)(d-2)^2(d-1)d} \bigg[ \big(-120+130 d-15 d^2-5 d^3\big) \frac{\R^3}{4(d-1)} + \big(30-55 d+15 d^2\big) \R \R^{ab}\R_{ab} \bigg]\\
&+\frac{L^5}{6(d-8)(d-6)(d-4)(d-2)^3(d-1)} \bigg[ \\
~~~~&\big(-61440 + 167168 d - 123136 d^2 + 38008 d^3 - 4514 d^4 + 63 d^5 +  21 d^6\big) \frac{\R^4}{16 d^2 (d-1)^2} \\
 &~~~~+ \big(-7680+28576 d-42456 d^2+26246 d^3-7803 d^4+1080 d^5-63 d^6\big) \frac{\R^2 \R^{ab} \R_{ab}}{2 d^2 (d-1)^2} \\
 &~~~~+ \big(2328-1634 d+375 d^2-19 d^3 \big) \R^a_b \R^b_c \R^c_d \R^d_a \bigg]\\
&+ \frac{L^7}{12(d-10)(d-8)(d-6)(d-4)(d-2)^4(d-1)^3d} \bigg[ \\
&~~~~\big(-2457600+14538752 d-30459904 d^2+27699584 d^3-12618640 d^4+3046820 d^5\\
&~~~~~~~~-374764 d^6+21195 d^7-216 d^8-27 d^9 \big) \frac{\R^5}{16d^2(d-1)^2} \\
&~~~~+ \big(-307200+2124544 d-5869824 d^2+8344576 d^3-6276084 d^4+2580076 d^5\\
&~~~~~~~~-587415 d^6+71301 d^7-4509 d^8+135 d^9\big) \frac{\R^3 \R^{ab}\R_{ab}}{2d^2(d-1)^2} \\
&~~~~+ \big(122496-527600 d+589236 d^2-278756 d^3+62727 d^4-6092 d^5+189 d^6\big) \R \R^a_b \R^b_c \R^c_d \R^d_a \bigg] \\
&+ \frac{L^9}{12(d-12)(d-10)(d-8)(d-6)(d-4)(d-2)^5(d-1)^3} \bigg[\\
&~~~~\big( -235929600+2521890816 d-10093428736 d^2+19583016960 d^3-20243482112 d^4\\
&~~~~~~~~+11975617920 d^5-4208281360 d^6+879183244 d^7-105396104 d^8+6849353 d^9\\
&~~~~~~~~-208013 d^{10}+1155 d^{11}+77 d^{12} \big) \frac{\R^6}{64d^4(d-1)^4} \\
&~~~~+ \big(-58982400+689455104 d-3247906816 d^2+8067641344 d^3-11660653312 d^4\\
&~~~~~~~~+10114976464 d^5-5371636812 d^6+1759074292 d^7-349958269 d^8+40538978 d^9\\
&~~~~~~~~-2588868 d^{10}+83050 d^{11}-1155 d^{12}\big) \frac{\R^4 \R^{ab}\R_{ab}}{16d^4(d-1)^4} \\
&~~~~+ \big(23519232-162132480 d+436222208 d^2-533662448 d^3+344089348 d^4-125624216 d^5\\
&~~~~~~~~+26237659 d^6-2995067 d^7+170609 d^8-3645 d^9\big) \frac{\R^2 \R^a_b \R^b_c \R^c_d \R^d_a}{4d^2(d-1)^2} \\
&~~~~+ \big(-2636928+4990512 d-3718820 d^2+1421092 d^3-293971 d^4+31944 d^5\\
&~~~~~~~~-1767 d^6+38 d^7\big) \R^a_b \R^b_c \R^c_d \R^d_e \R^e_f \R^f_a \bigg]
+ \ldots \Bigg)~,
}
\myeq{ \label{eqn:ll_ct2}
I_{m=3,\ct} &= \int d^d x \sqrt{h} \Bigg(
-\frac{6 (-4+d) (-3+d) (-2+d) (-1+d)}{L^5} + \frac{3 (-4+d) (-3+d)}{L^3}\R \\
&+ \frac{9(d-3)}{L(d-2)} \bigg[\frac{-d}{4(d-1)}\R^2 + \R^{ab}\R_{ab} \bigg] \\
&+ \frac{15(d-4)(d-3)L}{2(d-6)(d-2)^2(d-1)d} \bigg[(d^2+6d-8)\frac{\R^3}{4(d-1)} + (3d-2)\R\R^{ab}\R_{ab} \bigg]\\
&+ \frac{L^3}{4(d-8)(d-6)(d-2)^3(d-1)} \bigg[ \\
&~~~~\big(184320-496896 d+403200 d^2-135768 d^3+18474 d^4-315 d^5-105 d^6\big) \frac{\R^4}{16d^2(d-1)^2} \\
&~~~~+ \big(23040-85152 d+127992 d^2-86238 d^3+28455 d^4-4632 d^5+315 d^6\big) \frac{\R^2 \R^{ab}\R_{ab}}{2d^2(d-1)^2} \\
&~~~~+ \big(-3960+2346 d-243 d^2-33 d^3\big) \R^a_b \R^b_c \R^c_d \R^d_a \bigg]\\
&+\frac{L^5}{8(d-10)(d-8)(d-6)(d-2)^4(d-1)^3d} \bigg[ \big(7372800-39450624 d+79469568 d^2-72985728 d^3\\
&~~~~+34270704 d^4-8733180 d^5+1235700 d^6-92301 d^7+1512 d^8+189 d^9\big) \frac{\R^5}{16d^2(d-1)^2} \\
&~~~~+ \big(921600-5852928 d+15351168 d^2-21246912 d^3+16007436 d^4-6722340 d^5\\
&~~~~~~~~+1619457 d^6-233667 d^7+21051 d^8-945 d^9\big) \frac{\R^3 \R^{ab}\R_{ab}}{2d^2(d-1)^2} \\
&~~~~+ \big(-243072+1026960 d-1097772 d^2+448092 d^3-66753 d^4+156 d^5+429 d^6\big) \R  \R^a_b \R^b_c \R^c_d \R^d_a \bigg]\\
&+\frac{L^7}{8(d-12)(d-10)(d-8)(d-6)(d-2)^5(d-1)^3}\bigg[ \big(707788800-6401359872 d+23481679872 d^2\\
&~~~~~~~~-43466637312 d^3+43522156032 d^4-24779634048 d^5+8211306384 d^6-1587130092 d^7+180670920 d^8\\
&~~~~~~~~-13174929 d^9+644853 d^{10}-10395 d^{11}-693 d^{12}\big) \frac{\R^6}{64d^4(d-1)^4} \\
&~~~~+ \big(176947200-1777287168 d+7657439232 d^2-18007136256 d^3+25041792768 d^4-20965903440 d^5\\
&~~~~~~~~+10619859372 d^6-3234507732 d^7+582978213 d^8-62594418 d^9+4607652 d^{10}-280698 d^{11}\\
&~~~~~~~~+10395 d^{12}\big) \frac{\R^4 \R^{ab}\R_{ab}}{16d^4(d-1)^4} \\
&~~~~+ \big(-46669824+317514240 d-839054592 d^2+993035376 d^3-594270372 d^4+185533464 d^5\\
&~~~~~~~~-27917331 d^6+1259907 d^7+99783 d^8-8811 d^9\big) \frac{\R^2 \R^a_b \R^b_c \R^c_d \R^d_a}{4d^2(d-1)^2} \\
&~~~~+ \big(4599936-8142768 d+5263044 d^2-1517124 d^3+171147 d^4+1080 d^5-1161 d^6\\
&~~~~~~~~+66 d^7\big) \R^a_b \R^b_c \R^c_d \R^d_e \R^e_f \R^f_a \bigg] \Bigg)~.
}

\end{document}